\documentclass[twocolumn]{aastex701}
\usepackage{longtable}
\usepackage{multirow}
\usepackage{siunitx}
\usepackage{booktabs}
\usepackage{makecell}
\usepackage{amsmath}
\usepackage{enumitem}
\usepackage{subcaption}

\def \kms {\mathrm{km \, s^{-1}}}
\def \kpc {\mathrm{kpc}}
\def \kyr {\mathrm{kyr}}
\def \pc {\mathrm{pc}}
\def \Msun {M_\odot}

\def \modelA {$\Phi_{\mathrm{BH}}$}
\def \modelB {$\Phi_{\text{BH}} + \Phi_{\text{NSC}}(\mathrm{gas})$}
\def \modelC {$\Phi_{\text{BH}} + \Phi_{\text{NSC}}$}

\begin{document}

\title{Stellar-wind Fueled Accretion onto Sagittarius A* in the Presence of a Nuclear Star Cluster}

\author[0009-0005-5945-209X,sname='Skrabacz']{Edward Skrabacz}
\affiliation{Center for Interdisciplinary Exploration and Research in Astronomy, Northwestern University, Evanston IL 60208, USA}
\affiliation{Department of Physics \& Astronomy, Northwestern University, Evanston IL 60208, USA}
\email{edward.skrabacz@u.northwestern.edu}  

\author[0000-0001-8986-5403]{Lena Murchikova}
\affiliation{Center for Interdisciplinary Exploration and Research in Astronomy, Northwestern University, Evanston IL 60208, USA}
\affiliation{Department of Physics \& Astronomy, Northwestern University, Evanston IL 60208, USA}
\affiliation{School of Natural Sciences, Institute for Advanced Study, 1 Einstein Drive, Princeton, NJ 08540, USA}

\email{lena@northwestern.edu}  

\author[0000-0003-0220-5723]{Sean M. Ressler}
\affiliation{Canadian Institute for Theoretical Astrophysics, 60 St. George Street, Toronto, ON M5S 3H8, Canada}
\email{sressler@cita.utoronto.ca} 

\author[0009-0000-7650-7164]{Asad Ukani}
\affiliation{Center for Interdisciplinary Exploration and Research in Astronomy, Northwestern University, Evanston IL 60208, USA}
\affiliation{Department of Physics \& Astronomy, Northwestern University, Evanston IL 60208, USA}
\email{asadukani2028@u.northwestern.edu}

\author[0000-0001-6541-734X]{Siddhant Solanki}
\affiliation{Department of Astronomy, University of Maryland, 7901 Regents Drive, College Park, MD 20742, USA}
\email{siddhant@umd.edu}

\begin{abstract}

The Milky Way's Galactic Center hosts the black hole Sagittarius A* (Sgr A*), which provides us with a close-up view into supermassive black hole accretion and feedback. 
Recent works have shown that the winds from $\sim 30$ Wolf-Rayet (WR) stars orbiting Sgr A* at about 4 arcsec are important contributors to feeding the supermassive black hole.
A nuclear star cluster (NSC) with a mass of several $10^6 \, \text{M}_\odot$, of which $10^6 \, \text{M}_\odot$ is within 1 pc, also surrounds Sgr A*. 
The NSC contributes to the gravitational potential in the Galactic Center, affecting the orbits of the WR stars and their stellar winds. 
In this work, we examine the effects that the NSC has on the accretion of these stellar winds onto Sgr A* which have previously been neglected.
We find that, on the parsec scale, the effect from the gravitational potential of the NSC is negligible on the wind-fed accretion flow, validating the existing simulations used in the literature.

\end{abstract}

% where to search: https://astrothesaurus.org/concept-select/

\keywords{\uat{Accretion}{14} --- \uat{Active galactic nuclei}{16} --- \uat{Astrophysical black holes}{98} --- \uat{Galactic Center}{565} --- \uat{High Energy astrophysics}{739} --- \uat{Interstellar medium}{847} --- \uat{Magnetohydrodynamical simulations}{1966} --- \uat{Wolf-Rayet stars}{1806}}

\section{Introduction \label{sec: intro}} 

The center of the Milky Way Galaxy, commonly referred to as the Galactic Center, hosts a supermassive black hole named Sagittarius A* (Sgr A*). \
It has a mass of $\sim 4\times 10^6 \; M_\odot$ and is located $\sim 8$ kpc away from the Earth \citep{Gravity2019,Do2019}.
The inner few parsecs around Sgr A* are filled with several gaseous and stellar structures \citep{Genzel2010}. 
Among these are 31 Wolf-Rayet (WR) stars on orbits within $\sim 0.5$ pc that produce winds with temperatures $\geq 10^5$ K \citep{Paumard2006}. 
Previous studies have shown that these WR stars play an important role in feeding the black hole through their winds \citep{Cuadra2007,Ressler2018,Balakrishnan2024}, producing observed X-ray emission after shock heating to $\sim$ keV temperatures \citep{Baganoff2003}.

Additionally, the Galactic Center hosts a multitude of stars within the inner few parsecs of Sgr A* \citep{Schodel2009}.
The stars collectively form the nuclear star cluster (NSC), which has a mass of several $10^6 \, \text{M}_\odot$ to $10^7 \, \text{M}_\odot$, of which about $\approx10^6 \, \text{M}_\odot$ is within 1 pc \citep{Chatz2014,FK2017,Schodel2018}. 
The gravitational potential from the NSC becomes comparable to that of Sgr A* at a distance of $\sim1$ pc. The additional potential induces precession in the orbits of WR stars, affects the gas motion by hindering its ability to escape the Galactic Center, and can potentially alter the accretion rate onto Sgr A*.
Previous studies of the stellar wind accretion onto Sgr A* did not account for the gravitational potential of the NSC, nor how it influences gas and the stellar orbits.

In this work, we study the effects of the NSC gravitational potential on the wind-fed accretion flow onto Sgr A* in the inner $\sim1$ pc around the black hole using hydrodynamics simulations. 
We do this by improving on the previous simulations of the parsec-scale accretion flow onto Sgr A* (e.g., \citealt{Cuadra2007,Russell2017,Ressler2018,Ressler2019a, Ressler2019,Calderon2020,Calderon2025}) by adding the effects of the NSC potential on the winds and the stellar orbits. We track the properties of gas from parsec to sub-milliparsec scales, and compare them to previous works that did not include these effects.

This paper is organized as follows. In Section \ref{sec: methods}, we describe the simulation setup, including the implementation of the NSC potential, and the evolution of the precessing WR stars within the simulation. In Section \ref{sec: models} we present the three models we use to test the effects of the NSC.
In Section \ref{sec: results}, we present the results. We conclude in Section \ref{sec: conc}.

%%%%%%%%%%%%%%%%%%%%%%%%%%%%%%%%%%%%%%%%%%%
\section{Simulation Setup \label{sec: methods}}
%%%%%%%%%%%%%%%%%%%%%%%%%%%%%%%%%%%%%%%%%%%

%%%%%%%%%%%%%%%%%%%%%%%%%%%%%%%%%%%%%%%%%%%
\subsection{Fluid Equations and NSC potential}
%%%%%%%%%%%%%%%%%%%%%%%%%%%%%%%%%%%%%%%%%%%

Our simulations use the multi-purpose fluid dynamics code \verb|Athena++| \citep{White_2016, Athena++} and follow the Sgr A* wind-fed accretion setup of \cite{Ressler2018}. 
\verb|Athena++| is widely used throughout the literature as a fast and reliable magnetohydrodynamics (MHD) code that solves the equations of astrophysical fluid dynamics using finite volume methods.
For this work, we choose to adopt the Harten-Lax-van Leer-Einfeldt (\verb|HLLE|) Riemann solver \citep{HLLE} with a piecewise-linear reconstruction model.

Following \citet{Solanki2023}, we modify the hydrodynamic fluid equations to include stellar wind source terms, optically-thin cooling, and the gravitational effects of the NSC:
\begin{eqnarray} \label{eq: fluid}
    \frac{\partial \rho}{\partial t} &+& \nabla \cdot\left(\rho \textbf{v}\right) =  f\dot{\rho}_\text{w}
     \\
     \frac{\partial (\rho\textbf{v})}{\partial t} &+& \nabla \cdot\left(P \textbf{I} + \rho \textbf{vv}\right)  =   - \frac{\rho GM_\text{BH}}{r^2} \hat{r} - \rho \nabla \Phi_{\text{NSC}}\nonumber
     \\
     &&\quad  +  f \dot{\rho}_{\text{w}} \left<\textbf{v}_{\text{w,net}}\right> 
     \\
     \frac{\partial E}{\partial t}  &+&  \nabla \left[\left(E + P\right) \textbf{v}\right] =   \frac{\rho GM_\text{BH}}{r} \textbf{v}\cdot\hat{r} - Q_-
     \nonumber
     \\
      && \quad - \rho \textbf{v} \cdot \nabla \Phi_{\text{NSC}}  +\frac{1}{2} f \dot{\rho}_{\text{w}} \left<|\textbf{v}_{\text{w,net}}|^2\right> , \label{eq: fluid3}
\end{eqnarray}
where $\rho$ is the fluid mass density, \textbf{v} is the fluid velocity vector, $f$ is the fractional volume of a cell that a wind occupies, $\rho_\text{w}$ is the mass-loss density rate of the WR winds, $P$ is the fluid pressure, $GM_\text{BH}$ is the gravitational parameter of Sgr A*, $\hat{r}$ is the unit position vector from Sgr A*, $\textbf{v}_{\text{w}}$ is the velocity of the stellar winds, $Q_-$ is the cooling rate per unit volume, $E$ is the fluid energy density, and $\Phi_{\text{NSC}}$ is the gravitational potential of the NSC. 
Equations \ref{eq: fluid}-\ref{eq: fluid3} represent the conservation of mass, momentum, and energy, respectively. 

We use the NSC potential of \cite{Chatz2014}:
\begin{eqnarray}\label{eq: grad NSC}
    \nabla \Phi_{\text{NSC}} &=&
    - \frac{GM_1}{a_1} \left(\frac{1}{r + a_1} - \frac{r}{\left(r + a_1\right)^{2}}\right) \nonumber \\
    && \times \left(\frac{r}{r + a_1}\right)^{1 - \gamma_1},
\end{eqnarray}
where $M_1 = 2.7\times10^7 M_\odot$, $a_1 = 3.9$ pc, and $\gamma = 0.51$. 
We use this potential as it results in orbital evolutions consistent with observations and it is commonly used in the literature.
Note that the full NSC potential of \cite{Chatz2014} contains an additional term that becomes important only on the scale of $\sim 10$ pc to $100$ pc.
As we study only the inner $\sim 1$ pc, we neglect this additional term. 

We set the mass of Sgr A* to  $M_{\text{BH}} = 4.152 \times 10^6\ \Msun$ and its distance to be $8.178\ \kpc$ \citep{Gravity2019} from Earth, at which 1 arcsec $\simeq 0.04$ pc, consistent with previous simulations by \cite{Solanki2023} and \cite{Ressler2018}.
If we had instead used the estimate of \citet{Do2019}, which finds a mass of $M_{\text{BH}} = 3.964 \times 10^6\ \Msun$ and a distance of $7.946 \ \kpc,$ our results would be different by ${\sim}5\%,$ which would not quantitatively change our findings.

\subsection{Simulation Grid}
We perform our simulations on a  $(4 \text{ pc})^3$ Cartesian grid.
The simulations span a total time of $10$ kyrs, starting at 9 kyr in the past and ending 1 kyr into the future. We take $t=0$ in the simulations to reflect the present time.
The simulation grid contains seven layers of static mesh refinement (SMR), where the resolution doubles after each SMR layer.
Each level of mesh refinement has 128 cells in each direction, with boundaries at $\mp \, 2^{-(n + 1)} \text{ pc}$, where $n=1,...,7$ is the level of refinement, in each Cartesian direction. 

We apply different boundary conditions to the inner and outer edges of the simulation domain.
Gas reaching the inner boundary at radius $r_\text{in} \simeq 1200\; r_g\sim 0.2 \text{ mpc}$ from the center is removed from the simulation and considered accreted. 
Here, $r_g =GM_{\rm BH}/c^2$ is the gravitational radius of the black hole, and $c$ is the speed of light.
At the outer edges of the grid, we impose ``outflow'' boundary conditions, ensuring that material never re-enters the simulation if it reaches the edge of the box.

\subsection{Cooling}
The Galactic Center contains both hydrogen-rich and hydrogen-deficient gas at varying scales.
The temperature of this gas ranges from $10^2$ K to $10^9$ K, thus a robust cooling function is necessary for presenting accurate results.
The winds from the WR stars have temperatures $\gtrsim 10^5$~K, meaning that for our purposes, a cooling function which incorporates high-end temperature effects only.
However, this gas can cool to the lower temperatures. Thus, we follow \citet{Solanki2023} by using a piecewise cooling function that covers temperatures from $10^2$~K to $10^9$~K for our cooling function.
This cooling function combines the low-temperature effects of atomic line cooling, reverberational cooling, and molecular collisions as described in \citet{Koyama2002} with the high-temperature cooling effects of thermal brehmsstrahlung and line emission in collisional ionization equilibrium as described in \citet{Ressler2018}. 
The cooling function is thus physically motivated for the Galactic Center.

\subsection{Floors and Ceilings}
In solving the conservative hydrodynamic equations, the density and pressure values in the simulations can reach negative values.
Numerical floors are necessary for preventing this unphysical outcome.
We set the density floor to $1.0 \times 10^{-7} \Msun \,\pc^{-3}$ and the pressure floor to $1.0\times10^{-10} \Msun \,\pc^{-1} \, \kyr^{-2}$, both in the simulation's units.
If a cell reaches a pressure or density value below these floors, that cell's pressure or density are replaced with the floor values.
We also set temperature floor of 100 K, which acts as a density-dependent pressure floor, allowing us to remain on our cooling curve.

The simulation can occasionally result in unrealistically high sound speeds and gas velocities.
To ensure that these unrealistic speeds are never reached, we set a velocity ceiling of $1.3\times 10^5 \, \kms$, which is equal to ten times the free-fall speed at the inner boundary ($r_\mathrm{in} \simeq 0.2 \text{ mpc}$) of the simulation.
If any component of the velocity ever reaches values above this ceiling, we set that component to the ceiling value.
If the sound speed ever reaches a value above the ceiling, we reduce the pressure such that the sound speed matches the ceiling value.
In practice, the ceiling value is rarely reached aside from a few cells during the initialization of the simulations, which present unphysically high speeds when the stellar wind source terms are instantaneously activated. 
As the simulation's timesteps are based on the fastest velocity in a given time, the velocity ceiling also sets the minimum timestep for our simulations. 

\subsection{Evolving Wolf-Rayet Stars} \label{subsec: WR evol}

We include 31 of the observed WR stars in the inner $\sim$2 pc of the Galactic Center. Their mass ejection rates and orbital parameters are summarized in 
\cite{Paumard2006,Cuadra2007,Matins2007,YZ2015,Gillessen2017}.
Stars that do not have observationally determined orbital parameters are assigned orbital parameters by either minimizing the eccentricity of their orbits or (for stars inferred to be members of the clockwise stellar disk) ensuring that they remain in the disk \citep{Cuadra2007, Ressler2018}.
The positions of a few stars have been updated in \citet{von_Fellenberg_2022}, but these data were not used as they made a negligible difference in our initial positions and thus would not noticeably impact our results.

\begin{figure}
    \centering
    \includegraphics[width=0.45\textwidth]{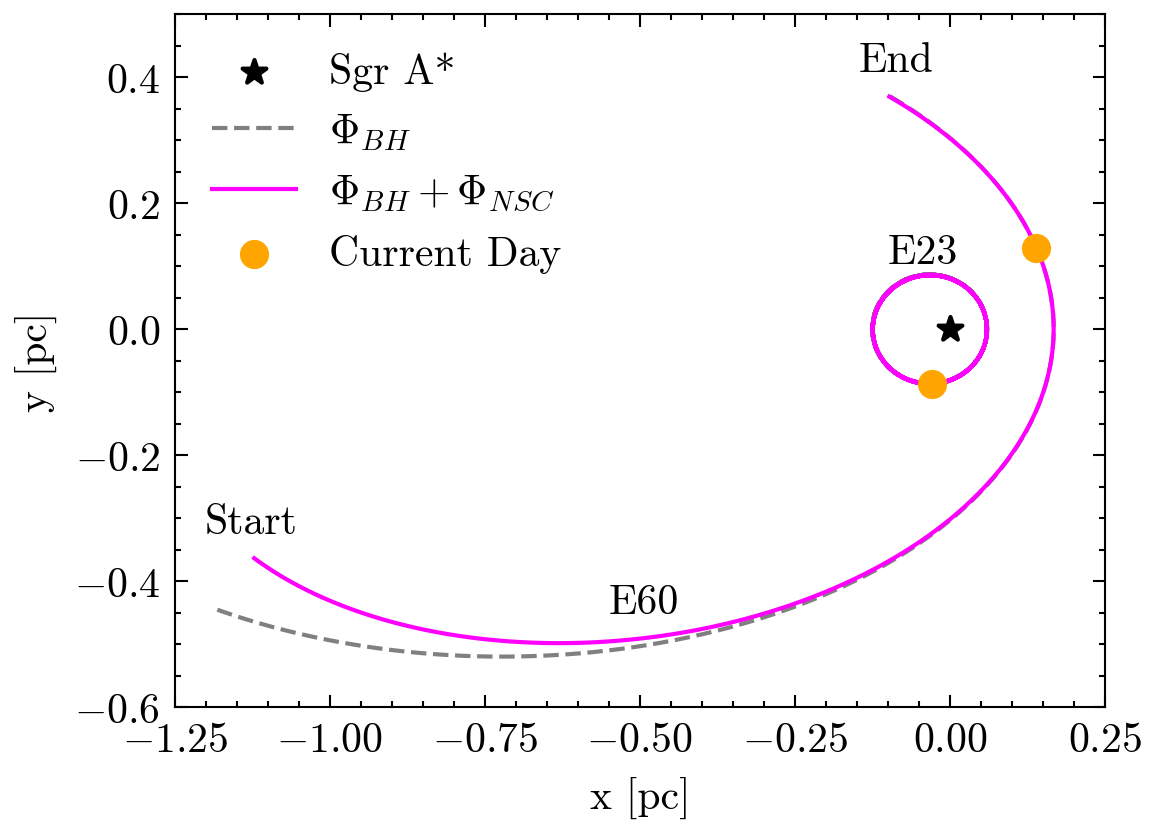}
    \caption{Comparison of WR star orbits under the influence of the gravitational potential of Sgr A* only ($\Phi_\mathrm{BH}$, dashed gray line) and the combination of the Sgr A* and NSC potentials ($\Phi_\mathrm{BH}+\Phi_\mathrm{NSC}$, solid magenta line). The orbits of two typical stars (E23 and E60 from \citealt{Paumard2006}) are plotted in their orbital planes with periapses directed towards the right. The yellow dots show the stars' current day position ($t=0$). Positions for the E60 star at the start and at the end of the simulation is marked. E23 star completes several orbits during the simulation so we only mark it's position at $t=0$.}
    \label{fig: WR Prec}
\end{figure}

\begin{figure*}[t!]
    \centering
    \includegraphics[width=0.97\linewidth]{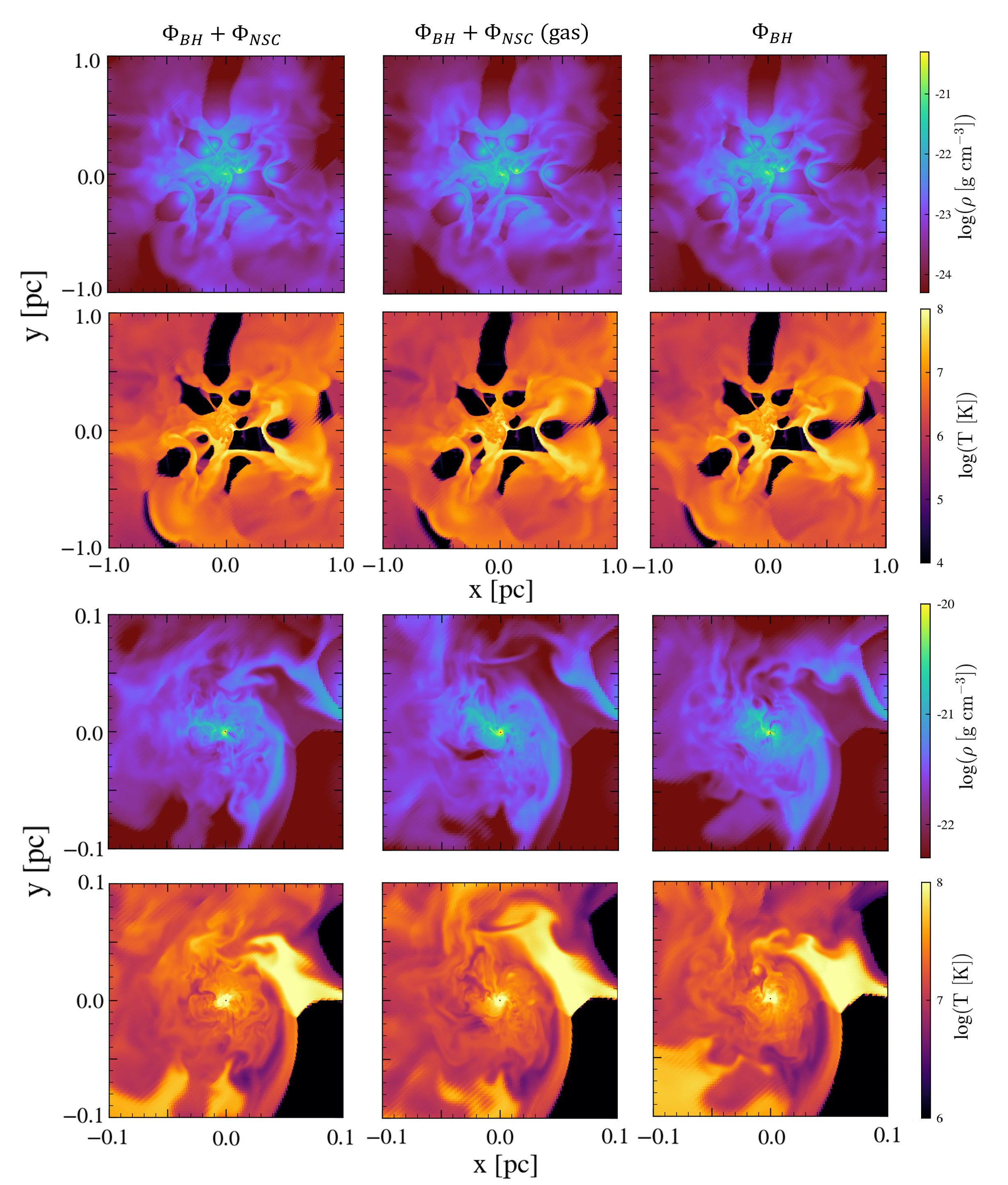}
    \caption{Comparison of the density and temperatures within 1 pc and 0.1 pc distance from Sgr A* for all three models. The name of the models are marked at the top. Details of the models are in Section \ref{sec: models}. The density and temperatures are calculated on the slice through the simulations volume passing through Sgr A* and parallel to the plane of the sky.  Sgr A* is positioned at the center of each plot. The difference in resolution in the all plots are due to the changes in the level of mesh refinement.}
    \label{fig: slice_full}
\end{figure*}

In the presence of the NSC, the stars no longer move on closed Keplerian orbits.
Since there is no closed form solution to the combined potentials from Sgr A* and NSC, we evolve the stars' positions and velocities using the Euler's method.
Compared to fourth-order Runge-Kutta (RK4), there is a negligible change in accuracy as the solutions converge for a timestep $dt \lesssim 10^{-4}$ yr.

The calculations were conducted in two distinct steps, which we name backward evolution and forward simulation.
The purpose of the two steps was to improve the speed of the simulation while ensuring that the orbits are accurately calculated.

\begin{figure*}[t!]
    \centering
    \includegraphics[width=1\linewidth]{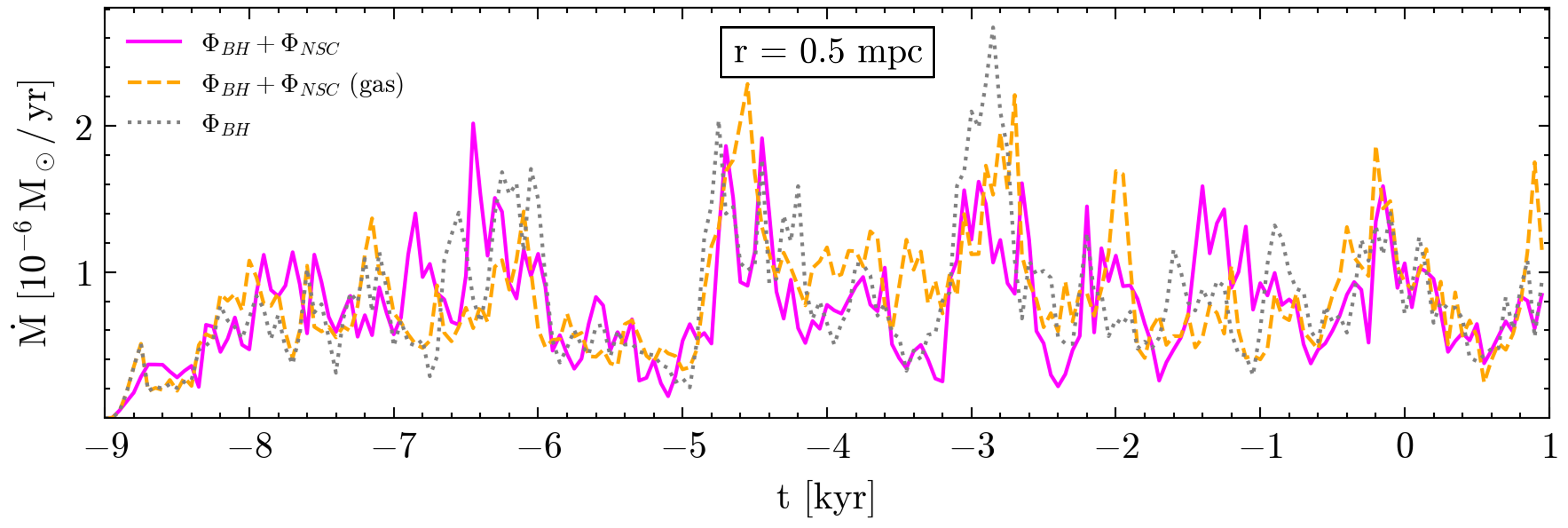}
    \caption{The accretion rate at r = $0.5$ mpc  from Sgr A* during the full duration of the simulation for all three models considered. The solid magenta line represents this accretion rate for \modelC, the dashed orange represents the accretion rate for \modelB, and the dotted grey represents the accretion rate for \modelA. The accretion rates are spatially averaged over solid angle and time-averaged over 50 years in simulation time. A detailed description of the models is given in Section \ref{sec: models}.} 
    \label{fig: mdot_overtime}
\end{figure*}

To ensure that the stars would arrive to the correct observed position at $t=0,$ we first evolve the WR stars 9 kyr backward from their current positions. 
We used the timestep $dt_{\text{evolve}} = 1\times 10^{-5}$ yr, which is $\sim 50$ times smaller than the smallest timestep in the simulation $(dt_\mathrm{sim})$. 
During this calculation, the positions and velocities of the WR stars are recorded to a table at every half year.
Saving at every half year allows for our table to be sufficiently small in memory and thus easily readable by the simulation.
Once the backward evolution reaches $t=-9$ kyr, the positions and velocities of the WR stars are recorded and used as starting points for their evolution in the simulations.

Starting at $t=-9$ kyr, the WR stars in the simulation are moved forward with RK1.
For this evolution with RK1, we use the varying $dt_{\text{sim}}$ of the simulation, which is calculated from the maximum velocity or sound speed ($c_s$) at a given time.
The relatively large and varying timesteps can introduce small errors which, if left unchecked, can cause the stars to incorrectly drift off their orbit.
To ensure that the WR stars arrive at their presently observed position and velocities, we check the stars' positions in the simulations against the tabulated values at every half year in simulation time.
If the difference between a star's simulation position and tabulated position is greater than the smallest grid in the simulation, then the star's simulated position is updated to the tabulated values.
By updating the position often, we keep the WR stars within their correct grid spaces.
Of the $\sim$540,000 such checks in the simulation, there are only around 500 instances of deviations across all stars, or, equivalently, a large enough deviation every $\sim 600$ years per star.
The first time a star's position was replaced happened at $t_{\text{sim}} \sim 400$ yr.

Figure \ref{fig: WR Prec} shows a comparison between two typical stellar orbits (stars E23 and E60). The black dashed line shows a Keplerian orbit which ignores the contribution of the NSC to the potential, and the solid magenta line is the full non-Keplerian orbit. Because Star E23 is so close to the black hole ($\lesssim 0.1$ pc), its motion is dominated by $\Phi_\mathrm{BH}$ and so the two orbits are indistinguishable. On the other hand, Star E60 shows clear deviation from the Keplerian trajectory.

The majority of WR stars' trajectories are nearly indistinguishable from Keplerian within 10 kyr of the simulations. Namely, their deviation from Keplerian trajectories is much smaller than the size of the the smallest simulation grid element.
However, around 10 stars with apoapses of $r_{\text{apo}} \geq 1$ pc do show major deviations.
The number of orbits completed by the stars in the simulations was between less than one and up to about 100.

\section{Models}\label{sec: models}

To determine the importance of the NSC for Sgr A*'s wind-fed accretion, we compare three models:

Model $\Phi_{\text{BH}}:$ The model which does not account for NSC potential. The stars and gas move in the gravitational potential of Sgr A* only. The orbits of the WR stars are strictly Keplerian.
    
Model $\Phi_{\text{BH}} + \Phi_{\text{NSC}}(\mathrm{gas}):$ The NSC's potential affects the motion of the gas in the simulation, but not the stars. The stars feel the gravitational effect of Sgr A* only, and their orbits of the WR stars are strictly Keplerian.
    
Model $\Phi_{\text{BH}} + \Phi_{\text{NSC}}:$ The NSC's potential affects the motion of both the gas and the stars. The orbits of the WR stars precess and deviate from Keplerian trajectories.

To implement these models in Equations \eqref{eq: fluid}-\eqref{eq: fluid3}, we set $\nabla\Phi_{\text{NSC}}  = 0$ for simulation $\Phi_{\text{BH}}$
and Equation \eqref{eq: grad NSC} for simulations $\Phi_{\text{BH}} + \Phi_{\text{NSC}}(\mathrm{gas})$ and $\Phi_{\text{BH}} + \Phi_{\text{NSC}}$. 

\section{Effect of the NSC on the Wind-fed Accretion Flow} \label{sec: results}

In this section, we evaluate the effect of the NSC on the wind-fed accretion by comparing the structure of the accretion flow in the three models described in Section \ref{sec: models}.

\begin{figure}
    \includegraphics[width=0.45\textwidth]{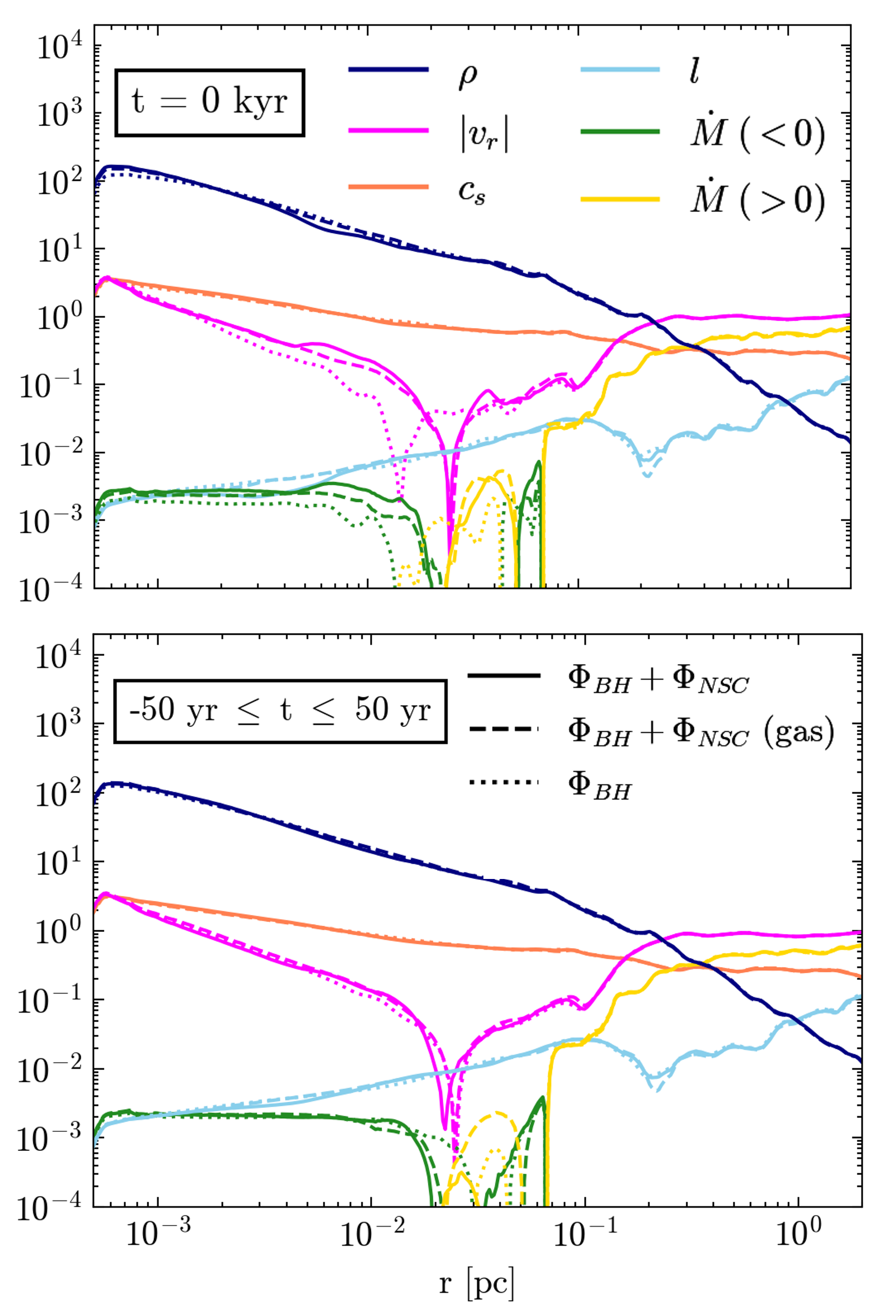}    
    \caption{Spatially-averaged radial profiles for simulations (a), (b), and (c) measured around the current day. The solid, dashed, and dotted lines represent said profiles for different hydrodynamic variables of simulations (c), (b), and (a) respectively. Purple lines represent mass density ($\rho$) in units $M_\odot/\text{pc}^3$, magenta lines represent radial velocity ($|v_r|$) in units $\text{pc}/\text{kyr}$, orange lines represent sound speed ($c_s$) in units $\text{pc}/\text{kyr}$, sky blue lines represent specific angular momentum ($l$) in units $M_\odot \, \text{pc}/\text{kyr}$, and the green and yellow lines represent the total accretion rate ($\dot{M}$) in units $M_\odot / \text{kyr}$. The green (yellow) lines show a net inwards (outwards) accretion flow. \textit{Top}: the radial profiles from the current day. \textit{Bottom}: The same radial profiles time averaged over $-50 \text{ yr}\leq t \leq 50 \text{ yr}$.}
   \label{fig: mdot_rad}
\end{figure}

Figure \ref{fig: slice_full} shows density and temperature slices of the gas in the plane of the sky for the three models in the inner 1 pc and 0.1 pc around the black hole taken at present day ($t=0$).
Although there are very subtle, visible differences, the three examined models produce nearly identical results.

Figure \ref{fig: mdot_overtime} shows the inwards accretion rate as a function of time for all models at a distance of $r = 0.5 \text{ mpc }\simeq 2 \,r_{\text{in}}$ from the black hole.
At early times, $-9 \, \kyr \leq t \leq -8 \, \kyr$, models \modelA \, and \modelB \, are in full agreement.
However, \modelC \, shows differences of order $10 \%$ and above in comparison to the other models at these early times.
The agreement between \modelA\, and \modelB \, over this interval is unsurprising, as the accretion in the beginning of the simulations is dominated by the winds from various stars in identical positions near the black hole (i.e. within 0.25 pc).
The gravitational potential is much weaker at this distance, so the winds from these stars remains unperturbed.
The divergence in the accretion rate in \modelC \, at these early times can be attributed to the differing positions in the stars as described by Figure \ref{fig: WR Prec}.
After this brief transient phase, the effects of the NSC gravitational potential become important as all models diverge from each other. 
However, the general structure of the accretion timeline is relatively similar across all models.
All three simulations tend to have peaks and troughs at similar times, with deviations mostly in the height of these features.
The accretion flow is similar between all models because it is determined by the positions of the stars, which differ only slightly.
 
To understand the present day accretion flow, we show radial profiles of a variety of hydrodynamic variables in Figure \ref{fig: mdot_rad}.
For the current day measurement (the top panel), the only noticeably large differences appear in the positions of the sharp dip in radial velocity $(|v_r|)$ between $10^{-2} \text{ pc} < r < 10^{-1}$ pc, which corresponds to a change between inflowing gas [$\dot{M} (< 0)$] and outflowing gas [$\dot{M} (> 0)$]. 
This switch corresponds to the ``stagnation region'', a region in which the outflowing gas effectively pushes away any inflowing gas, as defined in \citet{Ressler2018}. 
Notice that all three simulations contain this region between $0.01 \text{ pc}\lesssim r \lesssim 0.1 \text{ pc}$, matching the results from \citet{Ressler2018}.
The density ($\rho$), sound speed ($c_s$), and specific angular ($l$) are consistent across each model, following approximate $r^{-0.8}$, $r^{-0.5}$, and $r^{0.6}$ power laws, respectively, on the domain $10^{-3} \, \pc \leq r \leq 10^{-2} \, \pc$.
In this same domain, $|v_r| \propto r^{-1}$ and $\dot{M} \propto {r^{-0.2}}$.
All of these profiles match similar distributions from \citealt{Ressler2018,Ressler2019}.

Once we time-average (as seen in the bottom panel of Figure \ref{fig: mdot_rad}), nearly all differences between the simulations vanish. 
For instance, all the peaks of the $|v_r|$ profiles approach the same radial position.
The accretion rates still show slight differences in the previously mentioned stagnation region, however, the general structure remains relatively the same.
There is also no noticeable difference in the accretion rate structure near the inner and outer boundaries.

In comparison, other works which have included additional physics, such as magnetic fields, show far more dramatic differences.
For instance, in \cite{Ressler2019}, the addition of magnetization in the stellar winds causes the previously mentioned stagnation region to completely vanish.
Our simulations show no such strong differences between the three models.

\section{Conclusion \label{sec: conc}}
We have conducted simulation studies of the effect of the Galactic Center nuclear star cluster on stellar wind-fed accretion onto the Sgr A* black hole. We compared the structures of the accretion flow in three cases: the gas and the WR stars are affected by the gravity of the central black hole only (\modelA); the gas is affected by the gravity of the black hole and the NSC, but the WR stars by the black hole only and thus stay on the Keplerian orbits (\modelB); and both the gas and the stars are affected by the gravity of the black hole and the NSC (\modelB).

All three models present similar accretion rates in the inner regions of each simulation (Figure \ref{fig: slice_full}).
At early times, simulations \modelA \, and \modelB \, show the same trend, while \modelC \, differs between $10 \% - 100\%$ as a result of the difference in the WR stars' orbits.
The accretion rates of each model at later times converge to similar trends with minor differences.

Similar trends across models also seen in hydrodynamic variables at the present day, where we find that any differences can be time-averaged away over relatively short time-scales (Figure \ref{fig: mdot_rad}).
The differences present in the accretion flow parameters and the accretion rates can thus be seen as a result of the stochastic nature of the environment. No global differences between potentials with or without NSC is observed.

The inclusion of the NSC potential lead to such a small the change in the wind-red accretion flow parameters, that is even smaller than due to inclusion of the magnetic field which were previously deemed ``small''  \citep{Ressler2019}. 
We conclude that the gravitational potential of the Galactic Center NSC can be neglected when modeling the sub-parsec accretion flow onto Sgr~A*.

\section{Acknowledgments}

ES would like to thank Divjyot Singh for their discussions on numerical methodology along with Andrea Ceja, Nicole Flors, Pavani Jairam, Ved Shah, and Aswin Suresh for their advice and continuous encouragement. SMR is supported by the Natural Sciences and Engineering Research Council of Canada (NSERC), [funding reference number 568580]
Cette recherche a \'et\'e financ\'ee par le Conseil de recherches en sciences naturelles et en g\'enie du Canada (CRSNG), [num\'ero de r\'ef\'erence 568580].
SS is supported by the NSF grants AST-2307395 and AST-2406908 and the Simons Foundation grant MP-SCMPS-00001470.

This work used computing resources provided by Northwestern University and the Center for Interdisciplinary Exploration and Research in Astrophysics (CIERA). This research was supported in part through the computational resources and staff contributions provided for the Quest high performance computing facility at Northwestern University, which is jointly supported by the Office of the Provost, the Office for Research, and Northwestern University Information Technology.

\textit{Software:} \verb|Athena++| \citep{White_2016, Athena++}, \verb|yt| \citep{yt}

%\newpage
\bibliographystyle{aasjournalv7}
\bibliography{main}{}

@article{Ressler2018,
   title={Hydrodynamic simulations of the inner accretion flow of Sagittarius A* fuelled by stellar winds},
   author={{Ressler}, S M and {Quataert}, E and {Stone}, J M},
   year={2018},   
   volume={478},
   ISSN={1365-2966},
   url={http://dx.doi.org/10.1093/mnras/sty1146},
   doi={10.1093/mnras/sty1146},
   number={3},
   journal={Monthly Notices of the Royal Astronomical Society},
   publisher={Oxford University Press (OUP)},
   month=may, 
   pages={3544--3563} }

@article{Ressler2019,
   title={The surprisingly small impact of magnetic fields on the inner accretion flow of Sagittarius A* fueled by stellar winds},
   author={{Ressler}, S M and {Quataert}, E and {Stone}, J M},
   year={2019},
   volume={492},
   ISSN={1365-2966},
   url={http://dx.doi.org/10.1093/mnras/stz3605},
   doi={10.1093/mnras/stz3605},
   number={3},
   journal={Monthly Notices of the Royal Astronomical Society},
   publisher={Oxford University Press (OUP)},
   month=dec, 
   pages={3272--3293} }

@ARTICLE{Ressler2019a,
       author = {{Ressler}, S.~M. and {Quataert}, E. and {Stone}, J.~M.},
        title = "{Accretion of magnetized stellar winds in the Galactic centre: implications for Sgr A* and PSR J1745-2900}",
      journal = {Monthly Notices of the Royal Astronomical Society},
         year = 2019,
        month = jan,
       volume = {482},
       number = {1},
        pages = {L123-L128},
          doi = {10.1093/mnrasl/sly201},
archivePrefix = {arXiv},
       eprint = {1810.08617},
 primaryClass = {astro-ph.HE},
       adsurl = {https://ui.adsabs.harvard.edu/abs/2019MNRAS.482L.123R}
}

@article{Solanki2023,
   title={The Inner 2 pc of Sagittarius A*: Simulations of the Circumnuclear Disk and Multiphase Gas Accretion in the Galactic Center},
   author={{Solanki}, Siddhant and {Ressler}, Sean M. and {Murchikova}, Lena and {Stone}, James M. and {Morris}, Mark R.},
   year={2023},
   volume={953},
   ISSN={1538-4357},
   url={http://dx.doi.org/10.3847/1538-4357/acdb6f},
   doi={10.3847/1538-4357/acdb6f},
   number={1},
   journal={The Astrophysical Journal},
   publisher={American Astronomical Society},
   month=jul, 
   pages={22} }

@article{Cuadra2007,
   title={Variable accretion and emission from the stellar winds in the Galactic Centre},
   author={Cuadra, Jorge and Nayakshin, Sergei and Martins, Fabrice},
   year={2007},
   volume={383},
   ISSN={1365-2966},
   url={http://dx.doi.org/10.1111/j.1365-2966.2007.12573.x},
   doi={10.1111/j.1365-2966.2007.12573.x},
   number={2},
   journal={Monthly Notices of the Royal Astronomical Society},
   publisher={Oxford University Press (OUP)},
   month=dec, 
   pages={458--466} }

@article{Paumard2006,
   title={The Two Young Star Disks in the Central Parsec of the Galaxy: Properties, Dynamics, and Formation},
   author={Paumard, T. and Genzel, R. and Martins, F. and Nayakshin, S. and Beloborodov, A. M. and Levin, Y. and Trippe, S. and Eisenhauer, F. and Ott, T. and Gillessen, S. and Abuter, R. and Cuadra, J. and Alexander, T. and Sternberg, A.},
   year={2006},
   volume={643},
   ISSN={1538-4357},
   url={http://dx.doi.org/10.1086/503273},
   doi={10.1086/503273},
   number={2},
   journal={The Astrophysical Journal},
   publisher={American Astronomical Society},
   month=jun, 
   pages={1011--1035} }

@article{Gillessen2017,
   title={An Update on Monitoring Stellar Orbits in the Galactic Center},
   author={Gillessen, S. and Plewa, P. M. and Eisenhauer, F. and Sari, R. and Waisberg, I. and Habibi, M. and Pfuhl, O. and George, E. and Dexter, J. and Fellenberg, S. von and Ott, T. and Genzel, R.},
   year={2017},
   volume={837},
   ISSN={1538-4357},
   url={http://dx.doi.org/10.3847/1538-4357/aa5c41},
   doi={10.3847/1538-4357/aa5c41},
   number={1},
   journal={The Astrophysical Journal},
   publisher={American Astronomical Society},
   month=feb, 
   pages={30} }

@article{Gravity2019,
   title={A geometric distance measurement to the Galactic center black hole with 0.3\% uncertainty},
   author={Abuter, R. and Amorim, A. and Bauböck, M. and Berger, J. P. and Bonnet, H. and Brandner, W. and Clénet, Y. and Coudé du Foresto, V. and de Zeeuw, P. T. and Dexter, J. and Duvert, G. and Eckart, A. and Eisenhauer, F. and Förster Schreiber, N. M. and Garcia, P. and Gao, F. and Gendron, E. and Genzel, R. and Gerhard, O. and Gillessen, S. and Habibi, M. and Haubois, X. and Henning, T. and Hippler, S. and Horrobin, M. and Jiménez-Rosales, A. and Jocou, L. and Kervella, P. and Lacour, S. and Lapeyrère, V. and Le Bouquin, J.-B. and Léna, P. and Ott, T. and Paumard, T. and Perraut, K. and Perrin, G. and Pfuhl, O. and Rabien, S. and Rodriguez Coira, G. and Rousset, G. and Scheithauer, S. and Sternberg, A. and Straub, O. and Straubmeier, C. and Sturm, E. and Tacconi, L. J. and Vincent, F. and von Fellenberg, S. and Waisberg, I. and Widmann, F. and Wieprecht, E. and Wiezorrek, E. and Woillez, J. and Yazici, S.},
   year={2019},
   volume={625},
   ISSN={1432-0746},
   url={http://dx.doi.org/10.1051/0004-6361/201935656},
   doi={10.1051/0004-6361/201935656},
   journal={Astronomy \& Astrophysics},
   publisher={EDP Sciences},
   month=may, 
   pages={L10} }

@article{von_Fellenberg_2022,
   title={The Young Stars in the Galactic Center},
   volume={932},
   ISSN={2041-8213},
   url={http://dx.doi.org/10.3847/2041-8213/ac68ef},
   DOI={10.3847/2041-8213/ac68ef},
   number={1},
   journal={The Astrophysical Journal Letters},
   publisher={American Astronomical Society},
   author={von Fellenberg, Sebastiano D. and Gillessen, Stefan and Stadler, Julia and Bauböck, Michi and Genzel, Reinhard and de Zeeuw, Tim and Pfuhl, Oliver and Amaro Seoane, Pau and Drescher, Antonia and Eisenhauer, Frank and Habibi, Maryam and Ott, Thomas and Widmann, Felix and Young, Alice},
   year={2022},
   month=jun, pages={L6} }

@article{FK2017,
    author = {Feldmeier-Krause, A. and Zhu, L. and Neumayer, N. and van de Ven, G. and de Zeeuw, P. T. and Schödel, R.},
    title = {Triaxial orbit-based modelling of the Milky Way nuclear star cluster},
    journal = {Monthly Notices of the Royal Astronomical Society},
    volume = {466},
    number = {4},
    pages = {4040-4052},
    year = {2016},
    month = {12},
    issn = {0035-8711},
    doi = {10.1093/mnras/stw3377},
    url = {https://doi.org/10.1093/mnras/stw3377},
    eprint = {https://academic.oup.com/mnras/article-pdf/466/4/4040/10873094/stw3377.pdf},
}

@article{Chatz2014,
    author = {{Chatzopoulos}, S. and {Fritz}, T. K. and {Gerhard}, O. and {Gillessen}, S. and {Wegg}, C. and {Genzel}, R. and {Pfuhl}, O.},
    title = {The old nuclear star cluster in the Milky Way: dynamics, mass, statistical parallax, and black hole mass},
    journal = {Monthly Notices of the Royal Astronomical Society},
    volume = {447},
    number = {1},
    pages = {948-968},
    year = {2014},
    month = {12},
    issn = {0035-8711},
    doi = {10.1093/mnras/stu2452},
    url = {https://doi.org/10.1093/mnras/stu2452},
    eprint = {https://academic.oup.com/mnras/article-pdf/447/1/948/4933825/stu2452.pdf},
}

@article{Athena++,
   title={The Athena++ Adaptive Mesh Refinement Framework: Design and Magnetohydrodynamic Solvers},
   volume={249},
   ISSN={1538-4365},
   url={http://dx.doi.org/10.3847/1538-4365/ab929b},
   DOI={10.3847/1538-4365/ab929b},
   number={1},
   journal={The Astrophysical Journal Supplement Series},
   publisher={American Astronomical Society},
   author={{Stone}, James M. and {Tomida}, Kengo and {White}, Christopher J. and {Felker}, Kyle G.},
   year={2020},
   month=jun, pages={4} 
}

@ARTICLE{Do2019,
       author = {{Do}, Tuan and {Hees}, Aurelien and {Ghez}, Andrea and
         {Martinez}, Gregory D. and {Chu}, Devin S. and {Jia}, Siyao and
         {Sakai}, Shoko and {Lu}, Jessica R. and {Gautam}, Abhimat K. and
         {O'Neil}, Kelly Kosmo and {Becklin}, Eric E. and {Morris}, Mark R. and
         {Matthews}, Keith and {Nishiyama}, Shogo and {Campbell}, Randy and
         {Chappell}, Samantha and {Chen}, Zhuo and {Ciurlo}, Anna and
         {Dehghanfar}, Arezu and {Gallego-Cano}, Eulalia and
         {Kerzendorf}, Wolfgang E. and {Lyke}, James E. and {Naoz}, Smadar and
         {Saida}, Hiromi and {Sch{\"o}del}, Rainer and {Takahashi}, Masaaki and
         {Takamori}, Yohsuke and {Witzel}, Gunther and {Wizinowich}, Peter},
        title = "{Relativistic redshift of the star S0-2 orbiting the Galactic Center supermassive black hole}",
      journal = {Science},
         year = 2019,
        month = aug,
       volume = {365},
       number = {6454},
        pages = {664-668},
          doi = {10.1126/science.aav8137},
archivePrefix = {arXiv},
       eprint = {1907.10731},
 primaryClass = {astro-ph.GA},
       adsurl = {https://ui.adsabs.harvard.edu/abs/2019Sci...365..664D}
}

@ARTICLE{Genzel2010,
   author = {{Genzel}, R. and {Eisenhauer}, F. and {Gillessen}, S.},
    title = "{The Galactic Center massive black hole and nuclear star cluster}",
  journal = {Reviews of Modern Physics},
archivePrefix = "arXiv",
   eprint = {1006.0064},
     year = 2010,
    month = oct,
   volume = 82,
    pages = {3121-3195},
      doi = {10.1103/RevModPhys.82.3121},
   adsurl = {http://adsabs.harvard.edu/abs/2010RvMP...82.3121G}
}

@ARTICLE{Schodel2018,
       author = {{Sch{\"o}del}, R. and {Gallego-Cano}, E. and {Dong}, H. and {Nogueras-Lara}, F. and {Gallego-Calvente}, A.~T. and {Amaro-Seoane}, P. and {Baumgardt}, H.},
        title = "{The distribution of stars around the Milky Way's central black hole. II. Diffuse light from sub-giants and dwarfs}",
      journal = {\aap},
         year = 2018,
        month = jan,
       volume = {609},
          eid = {A27},
        pages = {A27},
          doi = {10.1051/0004-6361/201730452},
archivePrefix = {arXiv},
       eprint = {1701.03817},
 primaryClass = {astro-ph.GA},
       adsurl = {https://ui.adsabs.harvard.edu/abs/2018A&A...609A..27S}
}

@ARTICLE{Balakrishnan2024,
       author = {{Balakrishnan}, Mayura and {Russell}, Christopher M.~P. and {Corrales}, Lia and {Calder{\'o}n}, Diego and {Cuadra}, Jorge and {Haggard}, Daryl and {Markoff}, Sera and {Neilsen}, Joey and {Nowak}, Michael and {Wang}, Q. Daniel and {Baganoff}, Frederick},
        title = "{Multistructured Accretion Flow of Sgr A*. II. Signatures of a Cool Accretion Disk in Hydrodynamic Simulations of Stellar Winds}",
      journal = {The Astrophysical Journal},
         year = 2024,
        month = oct,
       volume = {974},
       number = {1},
          eid = {99},
        pages = {99},
          doi = {10.3847/1538-4357/ad6866},
archivePrefix = {arXiv},
       eprint = {2406.14631},
 primaryClass = {astro-ph.HE},
       adsurl = {https://ui.adsabs.harvard.edu/abs/2024ApJ...974...99B}
}

@ARTICLE{Russell2017,
       author = {{Russell}, Christopher M.~P. and {Wang}, Q. Daniel and {Cuadra}, Jorge},
        title = "{Modelling the thermal X-ray emission around the Galactic Centre from colliding Wolf-Rayet winds}",
      journal = {Monthly Notices of the Royal Astronomical Society},
         year = 2017,
        month = feb,
       volume = {464},
       number = {4},
        pages = {4958-4965},
          doi = {10.1093/mnras/stw2584},
archivePrefix = {arXiv},
       eprint = {1607.01562},
 primaryClass = {astro-ph.HE},
       adsurl = {https://ui.adsabs.harvard.edu/abs/2017MNRAS.464.4958R}
}

@ARTICLE{Calderon2020,
       author = {{Calder{\'o}n}, Diego and {Cuadra}, Jorge and {Schartmann}, Marc and {Burkert}, Andreas and {Russell}, Christopher M.~P.},
        title = "{Stellar Winds Pump the Heart of the Milky Way}",
      journal = {ApJL},
         year = 2020,
        month = jan,
       volume = {888},
       number = {1},
          eid = {L2},
        pages = {L2},
          doi = {10.3847/2041-8213/ab5e81},
archivePrefix = {arXiv},
       eprint = {1910.06976},
 primaryClass = {astro-ph.GA},
       adsurl = {https://ui.adsabs.harvard.edu/abs/2020ApJ...888L...2C}
}

@ARTICLE{Calderon2025,
       author = {{Calder{\'o}n}, Diego and {Cuadra}, Jorge and {Russell}, Christopher M.~P. and {Burkert}, Andreas and {Rosswog}, Stephan and {Balakrishnan}, Mayura},
        title = "{The formation and stability of a cold disc made out of stellar winds in the Galactic centre}",
      journal = {Astronomy \& Astrophysics},
         year = 2025,
        month = jan,
       volume = {693},
          eid = {A180},
        pages = {A180},
          doi = {10.1051/0004-6361/202452800},
archivePrefix = {arXiv},
       eprint = {2411.00100},
 primaryClass = {astro-ph.GA},
       adsurl = {https://ui.adsabs.harvard.edu/abs/2025A&A...693A.180C}
}

@article{HLLE,
author = {Einfeldt, Bernd},
title = {On Godunov-Type Methods for Gas Dynamics},
journal = {SIAM Journal on Numerical Analysis},
volume = {25},
number = {2},
pages = {294-318},
year = {1988},
doi = {10.1137/0725021},

URL = { 
    
        https://doi.org/10.1137/0725021
    
    

},
eprint = { 
    
        https://doi.org/10.1137/0725021
    
    

}
}

@ARTICLE{Matins2007,
       author = {{Martins}, F. and {Genzel}, R. and {Hillier}, D.~J. and {Eisenhauer}, F. and {Paumard}, T. and {Gillessen}, S. and {Ott}, T. and {Trippe}, S.},
        title = "{Stellar and wind properties of massive stars in the central parsec of the Galaxy}",
      journal = {\aap},
         year = 2007,
        month = jun,
       volume = {468},
       number = {1},
        pages = {233-254},
          doi = {10.1051/0004-6361:20066688},
archivePrefix = {arXiv},
       eprint = {astro-ph/0703211},
 primaryClass = {astro-ph},
       adsurl = {https://ui.adsabs.harvard.edu/abs/2007A&A...468..233M}
}

@ARTICLE{YZ2015,
       author = {{Yusef-Zadeh}, F. and {Bushouse}, H. and {Sch{\"o}del}, R. and {Wardle}, M. and {Cotton}, W. and {Roberts}, D.~A. and {Nogueras-Lara}, F. and {Gallego-Cano}, E.},
        title = "{Compact Radio Sources within 30' of Sgr A*: Proper Motions,Stellar Winds, and the Accretion Rate onto Sgr A*}",
      journal = {\apj},
         year = 2015,
        month = aug,
       volume = {809},
       number = {1},
          eid = {10},
        pages = {10},
          doi = {10.1088/0004-637X/809/1/10},
archivePrefix = {arXiv},
       eprint = {1506.07182},
 primaryClass = {astro-ph.GA},
       adsurl = {https://ui.adsabs.harvard.edu/abs/2015ApJ...809...10Y}
}

@ARTICLE{Baganoff2003,
       author = {{Baganoff}, F.~K. and {Maeda}, Y. and {Morris}, M. and {Bautz}, M.~W. and {Brandt}, W.~N. and {Cui}, W. and {Doty}, J.~P. and {Feigelson}, E.~D. and {Garmire}, G.~P. and {Pravdo}, S.~H. and {Ricker}, G.~R. and {Townsley}, L.~K.},
        title = "{Chandra X-Ray Spectroscopic Imaging of Sagittarius A* and the Central Parsec of the Galaxy}",
      journal = {\apj},
         year = 2003,
        month = jul,
       volume = {591},
       number = {2},
        pages = {891-915},
          doi = {10.1086/375145},
archivePrefix = {arXiv},
       eprint = {astro-ph/0102151},
 primaryClass = {astro-ph},
       adsurl = {https://ui.adsabs.harvard.edu/abs/2003ApJ...591..891B}
}

@ARTICLE{Koyama2002,
       author = {{Koyama}, Hiroshi and {Inutsuka}, Shu-ichiro},
        title = "{An Origin of Supersonic Motions in Interstellar Clouds}",
      journal = {\apjl},
         year = 2002,
        month = jan,
       volume = {564},
       number = {2},
        pages = {L97-L100},
          doi = {10.1086/338978},
archivePrefix = {arXiv},
       eprint = {astro-ph/0112420},
 primaryClass = {astro-ph},
       adsurl = {https://ui.adsabs.harvard.edu/abs/2002ApJ...564L..97K}
}

@ARTICLE{Schodel2009,
       author = {{Sch{\"o}del}, R. and {Merritt}, D. and {Eckart}, A.},
        title = "{The nuclear star cluster of the Milky Way: proper motions and mass}",
      journal = {\aap},
         year = 2009,
        month = jul,
       volume = {502},
       number = {1},
        pages = {91-111},
          doi = {10.1051/0004-6361/200810922},
archivePrefix = {arXiv},
       eprint = {0902.3892},
 primaryClass = {astro-ph.GA},
       adsurl = {https://ui.adsabs.harvard.edu/abs/2009A&A...502...91S}
}

@article{White_2016,
doi = {10.3847/0067-0049/225/2/22},
url = {https://doi.org/10.3847/0067-0049/225/2/22},
year = {2016},
month = {aug},
publisher = {The American Astronomical Society},
volume = {225},
number = {2},
pages = {22},
author = {White, Christopher J. and Stone, James M. and Gammie, Charles F.},
title = {AN EXTENSION OF THE ATHENA++ CODE FRAMEWORK FOR GRMHD BASED ON ADVANCED RIEMANN SOLVERS AND STAGGERED-MESH CONSTRAINED TRANSPORT},
journal = {The Astrophysical Journal Supplement Series}
}

@ARTICLE{yt,
       author = {{Turk}, Matthew J. and {Smith}, Britton D. and {Oishi}, Jeffrey S. and {Skory}, Stephen and {Skillman}, Samuel W. and {Abel}, Tom and {Norman}, Michael L.},
        title = "{yt: A Multi-code Analysis Toolkit for Astrophysical Simulation Data}",
      journal = {\apjs},
         year = 2011,
        month = jan,
       volume = {192},
       number = {1},
          eid = {9},
        pages = {9},
          doi = {10.1088/0067-0049/192/1/9},
archivePrefix = {arXiv},
       eprint = {1011.3514},
 primaryClass = {astro-ph.IM},
       adsurl = {https://ui.adsabs.harvard.edu/abs/2011ApJS..192....9T}
}

\end{document}